\documentstyle[12pt]{article}
\def\be{\begin{equation}}
\def\ee{\end{equation}}
\def\ba{\begin{array}}
\def\ea{\end{array}}
\def\bea{\begin{eqnarray}}
\def\eea{\end{eqnarray}}

\begin{document}

\title{Microscopic approach to the spectator matter fragmentation from 400 to 1000 AMeV}
\author{Yogesh K. Vermani and Rajeev K. Puri
\footnote{rkpuri@pu.ac.in} \\
Department of Physics, Panjab University, \\
Chandigarh-160014, India.}

\maketitle

\begin{abstract}

A study of multifragmentation of gold nuclei is reported at
incident energies of 400, 600 and 1000 MeV/nucleon using
microscopic theory. The present calculations are done within the
framework of quantum molecular dynamics (QMD) model. The
clusterization is performed with advanced sophisticated algorithm
namely \emph{simulated annealing clusterization algorithm} (SACA)
along with conventional spatial correlation method. A quantitative
comparison of mean multiplicity of intermediate mass fragments
with experimental findings  of ALADiN group gives excellent
agreement showing the ability of SACA method to reproduce the
fragment yields. It also emphasizes the importance of clustering
criterion in describing the fragmentation process within
semi-classical model.
\end{abstract}

\section{Introduction}
\label{intro} A highly excited  system formed in a nucleus-nucleus
collision, as a rule, is expected to break into several pieces
consisting of free nucleons, light charged particles (LCP's),
intermediate mass fragments (IMF's) as well as heavier residue.
This phenomenon of breaking of colliding nuclei into several
pieces is known as multifragmentation
\cite{don,kunde,li,aich,peas,tsang,zbiri}. Due to its complex
dynamics, mechanism behind this picture of 'explosive' break up
(into several entities) is not yet known completely.

At low incident energies, excitation energy available to the
system is very small. Therefore, larger impact of collisions is
needed to break the system into pieces of different sizes. In
other words, fruitful destruction is possible only for the central
collisions. On the other hand, mutual correlations among nucleons
are preserved in peripheral collisions, therefore, not much
deviation from the initial picture will be seen. In contrast,
excitation energy deposited in the system is very large at higher
incident energies. Therefore, central collisions break the matter
into very smaller pieces and rarely one sees intermediate mass
fragments or heavy mass fragments in these events. Maximum number
of IMF's can only be seen at semi-central impact parameters. Large
number of experiments have witnessed this trend of fragmentation
at various incident energies and impact parameters. This change in
the behavior of fragment pattern is also termed as a \emph{rise
and fall} in the multifragmentation \cite{tsang,schut,ogi,hub}.

As we go further towards higher incident energies, maximal of IMF
multiplicity starts shifting towards peripheral geometries. Such
trends have been found and reported in several recent experiments
of ALADiN - collaboration \cite{kunde,tsang}. In addition,
manyfold aspects of spectator matter fragmentation have also been
studied for the collision of $^{197}Au+ ^{197}Au$ on ALADiN set up
at incident energies varying between 150 and 600 MeV/nucleon.
Recently, INDRA experiments extended the energy domain covering
the incident energies between 40 and 150 MeV/nucleon \cite{lukas}.
The sole motivation for all these experiments was the fantastic
physics that may emerge from the disintegration of excited systems
leading to the expansion of matter to low densities. This onset of
multifragmentation and afterward transition to vaporization phase
has also been linked to the concept of liquid-gas phase transition
of nuclear matter \cite{zbiri,kreutz,poch}. Such critical behavior
is, however, reported to be influenced by the finite size effects
\cite{li,jaq}.

All these experimental studies characterize the evolution of
heavy-ion reaction from dominant multifragment-decay channel to
complete disassembly into light charge particles (LCP's) and free
nucleons sometimes also termed as 'vaporization'. The very recent
study by Puri and Kumar \cite{rkp} analyzed the $^{40}Ca+ ^{40}Ca
$ reaction for incident energies between 20 and 1000 MeV/nucleon
and over entire impact parameter range. They predicted a clear
rise and fall of multiplicity in incident energy and impact
parameter planes.

On the theoretical front, not much success has been reported to
reproduce the ALADiN experimental data \cite{tsang,blaich,das}.
Theoretical approaches which follow the evolution of target and
projectile to complete disassembly of nuclear matter needs
secondary algorithm to clusterize the phase space. Even
afterburners have also been employed to extract fragments. The
present study aims to check whether microscopic reaction models
can explain the universality reported by ALADiN group \cite{schut}
in spectator fragmentation or not. Molecular dynamical models QMD
\cite{aich} and QPD \cite{boal} were found to explain some of the
features of this experimental data \cite{tsang}. This questions
the validity of molecular dynamics models (MDM). The fallacy was
largely attributed to the lack of advanced secondary clustering
models \cite{barron,goss,dorso}. The clustering criterion is one
of the basic ingredients that may control the reaction mechanism
in semi-classical models like quantum molecular dynamics model.

Recently, a novel clusterization algorithm based on the energy
minimization criteria namely \emph{simulated annealing
clusterization algorithm} (SACA) was proposed \cite{goss}. As a
first attempt, results with this algorithm were quite promising
one \cite{goss}. In ref. \cite{goss}, $^{197}Au+ ^{197}Au$
reaction was studied at incident energy of 600 MeV/nucleon. Based
on the ALADiN results, there one assumed that fragment pattern
does not change above 400 MeV/nucleon. Therefore, it remains to be
seen whether QMD model can reproduce this universality feature or
not. We plan to address this situation in this letter. We apply
this algorithm to ALADiN data at incident energies of 400, 600 and
1000 MeV/nucleon in order to see whether our approach can explain
the rise and fall phenomenon and universal behavior in spectator
fragmentation at such higher incident energies. It is worth
mentioning that SACA algorithm has been robust against
experimental data at lower tail of incident energies. In our
earlier studies \cite{jai}, SACA method was reported to reproduce
the charge yields at incident energies between 25 and 200 AMeV. In
this analysis, O+Ag/Br reactions were taken \cite{jai}. In another
study, SACA method was tested against INDRA experimental data at
50 AMeV \cite{hirsc}. In this study, Xe+Sn reaction was subjected
to multifragmentation and various variables such as charge, proton
like and IMFs yields, angular distribution, average kinetic
energies \emph{etc.} were analyzed. SACA method explained all
these observables quite nicley, whereas conventional method failed
badly \cite{hirsc}. Due to the fact that interaction energy among
fragments is ignored, this approach of SACA can not be applied to
incident energies below above mentioned one. To study
fragmentation in Au+Au reaction, we followed nuclear collisions
within QMD model \cite{aich}. The phase space thus generated is
clusterized using advanced SACA method.

\section{SACA formalism}
\label{model} To generate the phase space of nucleons, we use
\emph{quantum molecular dynamics} (QMD) model. For the details of
the QMD model, reader is referred to refs.~\cite{aich,goss}. The
next essential step is to clusterize the phase space stored at
various time steps in each event. The extensively used approach
assumes the correlating nucleons to belong to same fragment if
their centers are closer than 4 fm \emph{i.e.} $\mid{\bf
r}_{\alpha}-{\bf r}_{\beta}\mid\leq 4 fm $. It may often lead to
wrong results if applied at higher densities and hence can't
address the time scale of multifragmentation. This approach is
labeled as minimum spanning tree (MST) algorithm.

In our latest approach, fragments are constructed based on the
energy correlations. The pre-clusters obtained with the MST method
are subjected to a binding energy condition \cite{goss,saca}:

\begin{eqnarray}
\zeta_{i}&=&\frac{1}{N_{f}}\sum_{\alpha=1}^{N_{f}}
\left[\sqrt{\left(\textbf{p}_{\alpha}-\textbf{P}_{N_{f}}^{cm}\right)
^{2}+m_{\alpha}^{2}}-m_{\alpha} \right. \nonumber \\
& & +\left.\frac{1}{2}\sum_{\beta \neq
\alpha}^{N_{f}}V_{\alpha\beta}
\left(\textbf{r}_{\alpha},\textbf{r}_{\beta}\right)\right]<E_{Bind},
\label{be}
\end{eqnarray}
with $E_{bind}$ = -4.0 MeV if $N_{f}\geq3$ and $E_{bind} = 0$
otherwise. In eq.~(\ref{be}), $N^{f}$ is the number of nucleons in
a fragment and $\textbf{P}_{N_{f}}^{cm}$ is the center-of-mass
momentum of the fragment. The requirement of a minimum binding
energy excludes the loosely bound fragments which will decay at
later stage.

To look for the most bound configuration (MBC), we start from a
random configuration which is chosen by dividing whole system into
few fragments. The energy of each cluster is calculated by summing
over all the nucleons present in that cluster using eq.
(\ref{be}). Note that we neglect the interaction between the
fragments. The total energy calculated in this way will differ
from the total energy of the system \cite{saca}.

Let the total energy of a configuration k be $E_{k}(=
{\sum_{i}}N_{f}\zeta_{i})$, where $N_{f}$ is the number of
nucleons in a fragment and $\zeta_{i}$ is the energy per nucleon
of that fragment. Suppose a new configuration $k^{'}$ (which is
obtained by (a)transferring a nucleon from randomly chosen
fragment to another fragment or by (b) setting a nucleon free, or
by (c) absorbing a free nucleon into a fragment) has a total
energy $E_{{k}^{'}}$. If the difference between the old and new
configuration $\Delta E (= E_{{k}^{'}}-E_{k})$ is negative, the
new configuration is always accepted. If not, the new
configuration $k^{'}$ may nevertheless be accepted with a
probability of $exp (-\Delta E/\upsilon)$, where $\upsilon$ is
called the control parameter. This procedure is known as
Metropolis algorithm. The control parameter is decreased in small
steps. This algorithm will yield eventually the most bound
configuration (MBC). Since this combination of a Metropolis
algorithm with slowly decreasing control parameter $\upsilon$ is
known as \emph{simulated annealing}, so our approach is dubbed as
\emph{simulated annealing clusterization algorithm} (SACA). For
more details, we refer the reader to ref.~\cite{saca}.

\section{Results}

For the present study, we use a soft equation of state (EoS) along
with standard energy-dependent \emph{n-n} cross section
\cite{cug}. The soft EoS has been advocated by many studies
\cite{peas,tsang,blaich,das,goss,mages}. The phase space is
generated and stored at many time steps and is then subjected to
the above mentioned clusterization procedures. To address the time
scale of multifragmentation of spectator matter, we employed SACA
method as well as spatial correlation method (\emph{i.e.} MST).

The density of environment is often correlated with the prediction
of breaking of nuclear matter into pieces. One can also look
density distribution in coordinate space to investigate the
formation of fragments. We here compute average density of system
as :
\begin{eqnarray}
\langle\rho\rangle&=&  \left \langle \frac{1}{A_{T}+A_{P}}
\sum_{i=1}^{A_{T}+A_{P}}\sum_{j>i}^{A_{T}+A_{P}}\frac{1}{(2\pi
L)^{3/2}} \right. \nonumber \\
 & & \left.\times e^{-({\bf{r}}_{i}(t)-{\bf{r}}_{j}(t))^{2}/2L} \right
\rangle,
\end{eqnarray}

with ${\bf{r}}_{i}$ and ${\bf{r}}_{j}$ being the position
coordinates of $i^{th}$ and $j^{th}$ nucleons. The Gaussian width
L is fixed with standard value of 1.08 $fm$.
%\begin{figure} [!h]
%\vskip -0.5 cm
%%\onefigure{fig1.ps}
%\includegraphics[scale=0.46, trim=10 0 10 0]{fig1.ps}% Here is how to import EPS art
%\vskip -0.8 cm \caption {Top panel: Time evolution of the average
%nucleon density $\langle \rho/\rho_{o} \rangle$ reached in
%$^{197}Au+^{197}Au$ collision. Bottom panel: The heaviest fragment
%$\langle A^{max} \rangle$ obtained with SACA and MST analysis as a
%function of time in $^{197}Au+^{197}Au$ collision.} \label{fig1}
%%\vskip -0.4 cm
%\end{figure}
Figure 1 (top panel) shows the time evolution of average nuclear
density $\langle\rho/\rho_{o}\rangle$ for Au+Au system at incident
energies of 400, 600 and 1000 MeV/nucleon and at an impact
parameter of 6 fm. The average nuclear density reaches its maximal
around 25 fm/c. This time domain also witnesses the maximum
collision rate and nuclear interactions which are going on between
among target and projectile nucleons. This maximal density shifts
towards later times as we go down the incident energies. The fine
point is that there is an insignificant change in the density
profile while enhancing the incident energy by the factor of 2.5
times \emph{i.e.} going from 400 to 1000 MeV/nucleon. At the final
stage of the reaction, we don't see any significant change with
the incident energy. The bottom panel of fig. 1 shows the time
evolution of the heaviest fragment $\langle A^{max}\rangle$ using
MST  and SACA techniques. The MST method gives one big cluster at
the time of maximum density, whereas one sees striking ability of
SACA method in identifying the heaviest fragment quite early when
violent phase of the reaction still continues. This suggests that
evolution of multifragmentation is an intricate process. In other
words, fragmentation starts at quite early stage when nucleons are
still interacting among themselves vigorously. The early
recognition of heaviest fragment $\langle A^{max} \rangle$ rules
out its formation out of the neck region. \emph{i.e.} geometrical
overlap between projectile and target. This suggests the emission
of $\langle A^{max} \rangle$ from the spectator region. Similar
trends of transition from the participant to spectator
fragmentation has also been observed and reported by
ALADiN-collaboration \cite{schut}. This finding also confronts the
common standpoint of thermal origin of fragments \emph{i.e.}
fragments are created after the thermalization sets in. Further
after violent phase of reaction is over (\emph{i.e.} after 60
fm/c), binding energy of all clusters in SACA method is greater
than $E_{Bind}$, the minimum binding energy needed to bind the
group of nucleons into cluster. Fragments after time 60 fm/c leave
the reaction zone without nucleon-nucleon correlations being
destroyed further. Hence fragment configuration obtained at the
earlier time can be compared with experimental data. Strikingly,
earlier detection of fragments (not shown here) at all incident
energies upto 1000 MeV/nucleon gives us possibility to look into
the \emph{n-n} interactions when nuclear matter is still hot and
dense. Further, one is also free from the problem of stability of
fragments. The failure of MST method to detect the fragments also
questions its validity at incident energies as high as 1000
MeV/nucleon. Simple correlations method fails to detect the
fragments even at these high excitation energies. The further rise
in $\langle A^{max} \rangle$ after 60 fm/c using SACA technique is
due to the reabsorption of surrounding light fragments by the
heavier fragments. We see that heavier $\langle A^{max} \rangle$
survive at smaller incident energies than at higher incident
energies.
%\begin{figure}[!h]
%\centering \vskip -1.2cm
%\includegraphics[scale=0.47, trim=20 0 18 0] {fig2.ps}
%% Here is how to import EPS art
%\vskip -1.0 cm \caption {The mean multiplicity of intermediate
%mass fragments $\langle N_{IMF} \rangle$ as a function of impact
%parameter b for the reaction of $^{197}Au+^{197}Au$. The model
%calculations with SACA (solid squares) and MST (open triangles)
%methods are compared with experimental data (open circles)
%reported by ALADiN group \cite{schut}.} \label{fig2}
%\end{figure}
The capability of QMD model clubbed with SACA method is
illustrated in fig. 2 where we display the mean multiplicity of
intermediate mass fragments $\langle N_{IMF} \rangle$ as a
function of impact parameter of the reaction. Also shown are the
results obtained with MST method. Our model calculations with SACA
method are in close agreement with ALADiN data \cite{schut} for $
^{197}Au+ ^{197}Au $ reaction at all incident energies 400, 600
and 1000 MeV/nucleon. As seen in the fig. 2, we also achieved a
reasonable reproduction of the shape of impact parameter
dependence of $\langle N_{IMF} \rangle$. Due to shallow minima
sometimes, we also see second minima before 60 fm/c in peripheral
collisions. We show also the calculations at these minima marked
as (*). We see that fragment structure at these minima is further
closer to the data. Further the peak value of $\langle N_{IMF}
\rangle$ and the corresponding impact parameter b is also well
estimated with QMD + SACA method. The prominent feature of the
spectator decay is the invariant nature of the IMF distribution
with respect to the bombarding energy. The SACA method
successfully reproduced the universal nature of spectator
fragmentation at all the three bombarding energies. It is worth
interesting to note that these universal features observed in
multifragmentation of gold nuclei persist upto much higher
bombarding energies than explored in this work \cite{adam}. In
contrary, the normal spatial correlation method fails badly to
explain the production of intermediate mass fragments at all
incident energies. This questions the validity of MST method in
explaining the fragmentation pattern in heavy-ion collisions.

\section {Summary}
We have studied multifragment-emission in $^{197}Au+^{197}Au$
reaction at incident energies of 400, 600 and 1000 MeV/nucleon
where ALADiN experiments showed the universality in the production
of intermediate mass fragments. For this study, we employed QMD
model clubbed with energy minimization algorithm (SACA) along with
conventional spatial correlation method. Our findings reveal that
SACA is able to reproduce the universal nature of
multifragmentation of excited spectator over entire impact
parameter-energy plane whereas spatial correlation method failed
to reproduce the IMF multiplicity. This is for the first time that
QMD + SACA approach is able to  reproduce the entire energy
domain. It also shows that mass and multiplicity of spectator
fragments remain invariant to range of bombarding energies. This
also resolved the earlier discrepancy where QMD model
underestimated the fragment yield \cite{tsang,blaich} at large
impact parameters even after 200 fm/c. In our case, SACA method is
successful in breaking the spectator matter into intermediate mass
fragments. Our results show that the QMD model contains necessary
ingredients to describe the physics of spectator decay. The
clustering algorithm one uses, however, holds the key tenet to
explain the reaction mechanism. \\

\section{\label{ack} Acknowledgement}
This work was supported by a research grant from Department of
Science and Technology, Government of India. \\

%\newpage
%
{\Large \bf Figure Captions} \\

{\bf FIG. 1.} Top panel: Time evolution of the average nucleon
density $\langle \rho/\rho_{o} \rangle$ reached in
$^{197}Au+^{197}Au$ collision. Bottom panel: The heaviest fragment
$\langle A^{max} \rangle$ obtained with SACA and MST analysis as a
function of time in $^{197}Au+^{197}Au$ collision. \\

{\bf FIG. 2.}The mean multiplicity of intermediate mass fragments
$\langle N_{IMF} \rangle$ as a function of impact parameter b for
the reaction of $^{197}Au+^{197}Au$. The model calculations with
SACA (solid squares) and MST (open triangles) methods are compared
with experimental data (open circles) reported by ALADiN group
\cite{schut}. \\


\begin{thebibliography}{100}

\bibitem{don} Donangelo R. and Souza S. R., Phys. Rev. C 58(1998) R2659.

\bibitem{kunde}
Kunde G. J. {\it et al.}, Phys. Rev. Lett. 74 (1995) 38.

\bibitem{li} Li. T {\it et al.}, Phys. Rev. Lett. 70 (1998) 1924.

\bibitem{aich}Aichelin J., Phys. Rep.202(1991)233; ichelin J. and
Stoecker H., Phys. Lett. B 176 (1986)74.

\bibitem{peas}Peaslee G. F. {\it et al.}, Phys. Rev. C 49 (1994) R2271.

\bibitem{tsang} Tsang M. B. {\it et al.}, Phys. Rev. Lett. 71 (1993)1502.

\bibitem{zbiri}Zbiri K. {\it et al.}, Phys. Rev. C 75 (2007) 034612.

\bibitem{schut} Sch\"{u}ttauf A. {\it et al.}, Nucl .Phys. A 607 (1996)457.

\bibitem{ogi} Ogilvie C. A. {\it et al.}, Phys. Rev. Lett. 67 (1991)1214.

\bibitem{hub} Hubele J. {\it et al.}, Z. Phys. A 340 (1991) 263.

\bibitem{lukas} Lukasik J. {\it et al.}, Phys. Lett. B 566 (2003) 76.

\bibitem{kreutz} Kreutz P. {\it et al.}, Nucl. Phys. A 556 (1993)
672.

\bibitem{poch} Pochodzalla J. {\it et al.},
Phys. Rev. Lett. 75 (1995) 1040.
%
\bibitem{jaq} Jaqaman H. R., Mekjian A. Z. and Zamick L., Phys. Rev. C 29 (1984) 2067.

\bibitem{rkp} Puri R. K. and Kumar S., Phys. Rev. C 57 (1998) 2744.

\bibitem{blaich} Begemann-Blaich M. {\it et al.}, Phys. Rev. C 48 (1993) 610.

\bibitem{das} Das C. B., Das A., Satpathy M. and Satpathy L., Phys. Rev. C 56 (1997) 1444.

\bibitem{boal} Boal D. H. and Glosli J. N., Phys. Rev. C 38 (1988) 1870.

\bibitem{barron} Barra\~{n}\'{o}n A., L\'{o}pez J. A. and
Escamilla Roa J.,  Phys. Rev. C 69 (2004) 014601; Chernomoretz A.
and Dorso C. O., Eur. Phys. J. D 24 (1984) 197; Campi X., Phys.
Lett. B 208 (1998) 351.

\bibitem{goss} Gossiaux P. B., Puri R. K., Hartnack Ch. and
Aichelin J., Nucl. Phys. A 619 (1997) 379.

\bibitem{dorso} Dorso C. O. and Randrup J., Phys. Lett. B 301 (1993) 328.


\bibitem{jai} Singh J. and Puri R. K., J. Phys. G: Nucl. Part.
Phys. 27 (2001) 2091; Puri R. K., Singh J. and Kumar S., Pramana
J. Phys. 59 (2002) 19.

\bibitem{hirsc} Nebauer R., Guertin A., Puri R., Hartnack Ch., Gossiaux P. B. and Aichelin J. in
Proceedings of the International Workshop, Hirschegg, Austria, Ed.
Feldmeier H., Knoll J., Noerenberg W., and Wambach J., Vol. XXVII,
(1999) p.43-61.


\bibitem{saca} Puri R. K., Hartnack C. and Aichelin J., Phys. Rev. C 54 (1996) R28; Kumar S. and Puri R.
K., Phys. Rev. C 58 (1998) 320; Puri R. K. and Aichelin J., J.
Comput. Phys. 162 (2000) 245; Dhawan J. K. and Puri R. K. Phys.
Rev. C 75 (2007) 057601.
%
\bibitem{cug} Cugnon J., Mizutani T. and Vandermeulen J., Nucl. Phys. A 352 (1981) 505.

\bibitem{mages} Magestro D. J., Bauer W. and Westfall G. D.,
Phys. Rev. C 62 (2000) 041603(R).

\bibitem{adam} M. I. Adamovich {\it et al.}, Z. Phys. A 359 (1997) 277.

\end{thebibliography}
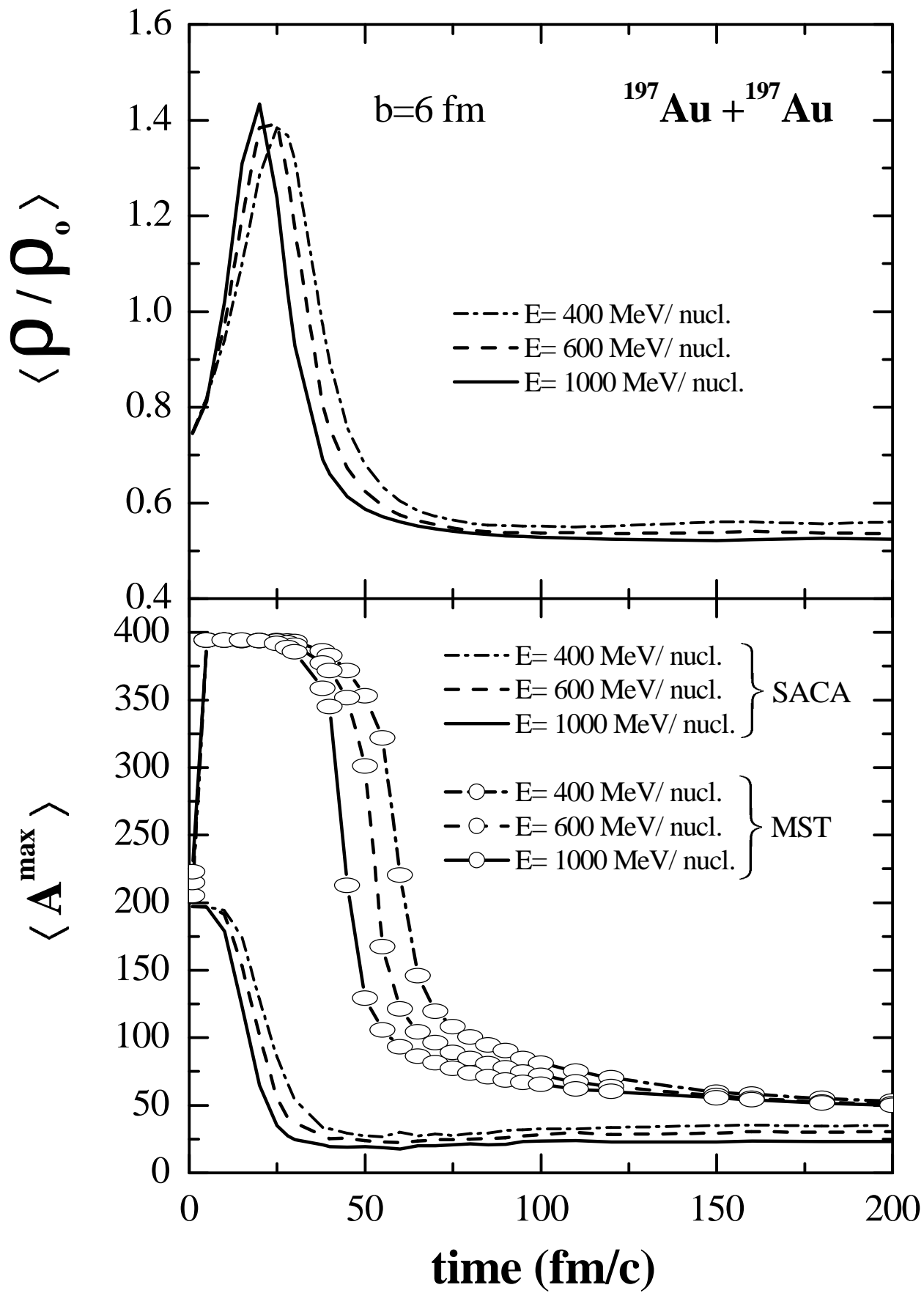
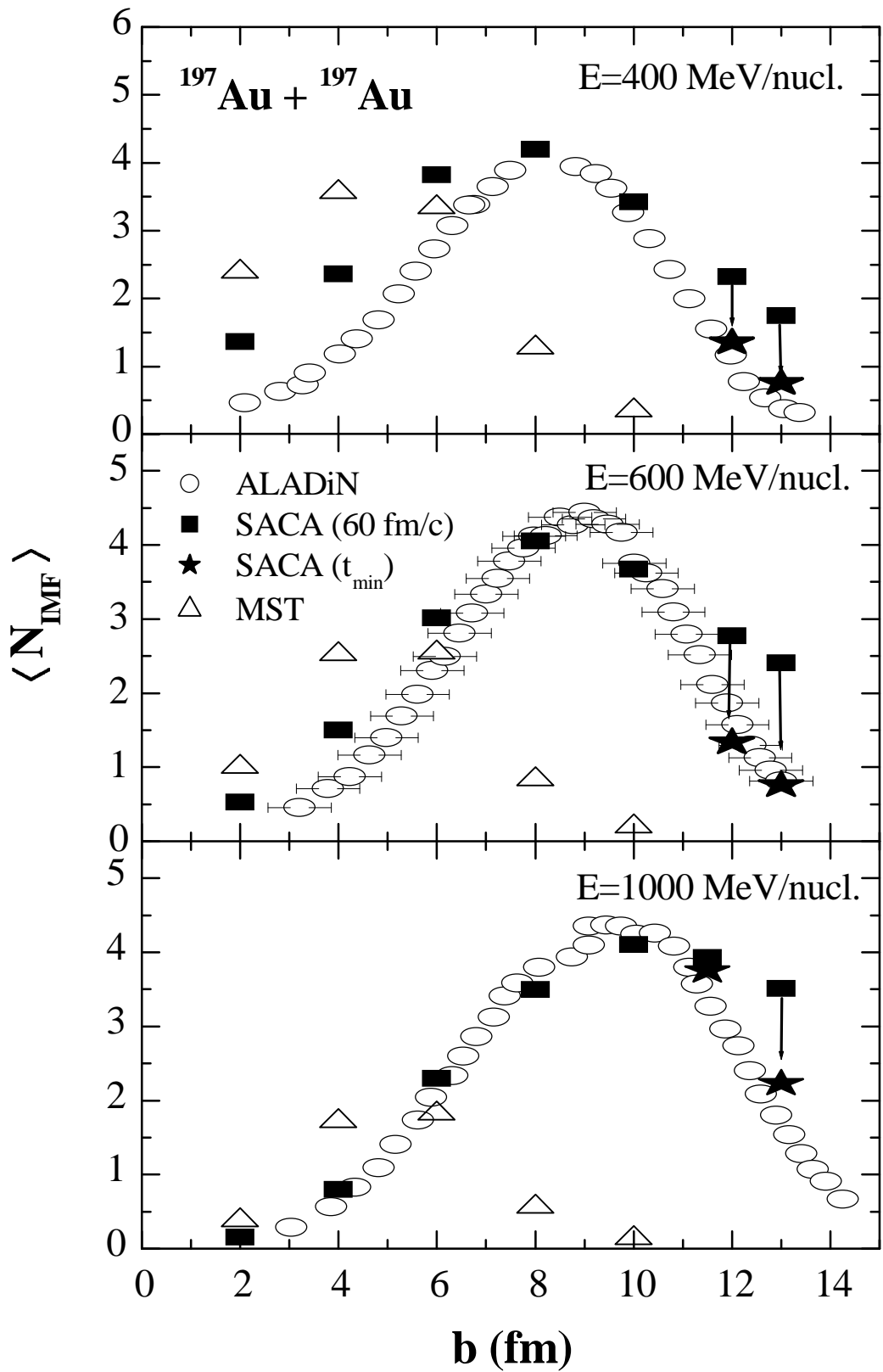
\end{document}